\documentclass[conference]{IEEEtran}

\usepackage[utf8]{inputenc}
\usepackage{cite}
\usepackage[ruled,vlined]{algorithm2e}
\usepackage[T1]{fontenc}
\usepackage{microtype}
\usepackage{graphicx}
\usepackage{paralist}
\usepackage{tabularx}
\usepackage{balance}
\usepackage{float}
\usepackage{multicol}
\usepackage{multirow}
\usepackage{pbox}
\usepackage{amssymb}
\usepackage{colortbl}
\usepackage{pifont}
\usepackage{xspace}
\usepackage[dvipsnames,table,xcdraw]{xcolor}
\usepackage{url}
\usepackage[TABBOTCAP]{subfigure}
\usepackage{rotating}
\usepackage{tikz}
\usepackage[normalem]{ulem}

\usepackage{enumitem}
\useunder{\uline}{\ul}{}

\usepackage{booktabs}

\newcommand{\ie}{\textit{i.e.,}\xspace}
\newcommand{\aka}{\textit{a.k.a.,}\xspace}
\newcommand{\eg}{\textit{e.g.,}\xspace}

\newcommand{\etal}{\textit{et al.}\xspace}

\newcommand{\secref}[1]{Section~\ref{#1}\xspace}

\newcommand{\figref}[1]{Fig.~\ref{#1}\xspace}

  {\list{}{\leftmargin=0.1in\rightmargin=0.1in}\item[]}%
  {\endlist}

\include{macro}

\begin{document}

\title{Continuous, Evolutionary and Large-Scale: A New Perspective for
	Automated Mobile App Testing}

\author{\IEEEauthorblockN{Mario Linares-V\'asquez$^1$,  Kevin Moran$^2$,  and Denys Poshyvanyk$^2$}
	\IEEEauthorblockA{%Department of Computer Science\\
		$^1$Universidad de los Andes, Bogot\'a, Colombia\\
		$^2$College of William \& Mary, Williamsburg, VA, USA%\\
		%m.linaresv@uniandes.edu.co,  \{kpmoran, denys\}@cs.wm.edu
	}}

\maketitle

\begin{abstract}
Mobile app development involves a unique set of challenges including device fragmentation and rapidly evolving platforms, making testing a difficult task. The design space for a comprehensive mobile testing strategy includes features, inputs, potential contextual app states, and large combinations of devices and underlying platforms. Therefore, automated testing is an \textit{essential} activity of the development process. However, current state of the art of automated testing tools for mobile apps poses limitations that has driven a preference for manual testing in practice. As of today, there is no comprehensive automated solution for mobile testing that overcomes fundamental issues such as automated oracles, history awareness in test cases, or automated evolution of test cases.

In this perspective paper we survey the current state of the art in terms of the frameworks, tools, and services available to developers to aid in mobile testing, highlighting present shortcomings. Next, we provide commentary on current key challenges that restrict the possibility of a comprehensive, effective, and practical automated testing solution.  Finally, we offer our vision of a comprehensive mobile app testing framework, complete with research agenda, that is succinctly summarized along three principles: Continuous, Evolutionary and Large-scale (CEL).

\end{abstract}
\IEEEpeerreviewmaketitle

\section{Introduction}
\label{sec:intro}
% !TEX root = main.tex

	The mobile handset industry has been growing at an unprecedented rate and the global ``app" economy, made up of millions of apps and  developers, and billions of devices and users, has been a tremendous success. This burgeoning mobile app market is fueled by rapidly evolving performant hardware and software platforms that support increasingly complex functionality. Currently, many modern mobile apps have practically the same features as their desktop counterparts and range in nature from games to medical apps. These mobile platforms enable user interaction via touch-screens and diverse sensors (e.g., accelerometer, temperature, gyroscope) that present new challenges for software testing.
	
	Unique characteristics and emerging best practices for creating mobile apps, combined with immense market interest, have driven both researchers and industrial practitioners to devise frameworks, tools, and services aimed at supporting mobile testing with the goal of assuring the quality of mobile apps.  However, current limitations in both manual and automated solutions underlie a broad set of challenges that prevent the realization of a comprehensive, effective, and practical automated testing approach\cite{Mona:ESEM13,Shauvik:2015,Kochhar:ICST15}.  Because of this, mobile app testing is still performed mostly manually costing developers, and the industry, significant amounts of effort, time, and money\cite{Mona:ESEM13,Shauvik:2015,Kochhar:ICST15}.  As development workflows increasingly trend toward adoption of agile practices, and continuous integration is adopted by larger numbers of engineers and development teams, it is imperative that automated mobile testing be enabled within this context if the development of mobile apps is to continue to thrive. However, current solutions for automated mobile testing do not provide a ``fully" automated experience, and several challenges are still open issues requiring attention from the community, if the expected goal is to help mobile developers to assure quality of their apps under specific conditions such as pressure from the users for continuous delivery and restricted budgets for testing processes.
	  
	 In this perspectives paper we present a new take on mobile app testing called CEL testing, which is founded on three principles: \emph{Continuous}, \emph{Evolutionary}, and \emph{Large-scale} (CEL). To properly frame and illustrate our vision, our paper (i) surveys the current state of research and practice for mobile testing, (ii) summarizes the key challenges deterring the creation of a comprehensive automated solution, (iii) offers a vision, informed by years of research, industrial collaborations, conversations with industry experts, for a comprehensive, practical automated testing solution, and (iv) outlines an actionable research agenda enabling this vision that we hope will be readily adopted by the SE research community.

\section{State-of-the-Art and Practice}
\label{sec:state-art}
% !TEX root = main.tex

	In order to adequately describe our vision for the future of mobile testing practices, it is first important to take a step back, and survey the state of research and practice.  
Thus, in this section, we give an overview of the frameworks, tools, and services that are currently available to support mobile application testing, hinting at current limitations that must be overcome to enable our proposed vision. In order to provide an ``at-a-gance" overview of the current state of mobile testing, we summarize solutions currently available to developers (see Table \ref{tab:android-testing-overview}).  Due to space limitations, we focus on the use cases and existing problems and challenges with the state of the art.

	It should be noted that a comprehensive overview of all available research and commercial software related to mobile testing is outside the scope of this paper and that our overview is by no means exhaustive.  The goal of presenting this information is to inform readers who may not be familiar with topic of mobile testing and motivate our vision by highlighting the current shortcomings of research and tools as informed by our past experience.  Thus, we limit our analysis to research generally concerned with \textit{functional} testing of mobile applications, and to popular commercial testing services and tools as gleaned from our previous research experience and industrial collaborations.  The 7 categories of tools presented were derived in different ways.  The first three categories (Automation Frameworks \& APIs, Record \& Replay Tools, and Automated Input Generation tools) have generally been defined by prior work \cite{Moran:ICST16,Shauvik:2015,Mao:ISSTA16}, and we expand upon these past categorizations.  The other four categories were derived by examining commercial software and service offerings available to mobile developers, as informed from our past experience.  We delineated the features of these offerings, and it was clear that some tools shared common dimensions, thus forming the categories we present in this section.

%-----------------------------------------------------------------
\subsection{Automation APIs/Frameworks}
\label{subsec:frameworks}

	One of the most basic, yet most powerful testing tools available to developers on several mobile platforms are the \textit{GUI-Automation Frameworks} and \textit{APIs} \cite{uiautomator,espresso,uiautomation,appium,roboelectric,calabash,cucumber,ranorex,robotium}.  These tools often serve as interfaces for obtaining GUI-related information such as the hierarchy of components/widgets that exist on a screen and for simulating user interactions with a device.  Because these frameworks and APIs provide a somewhat \textit{universal interface} to the GUI or underlying system functionality of a mobile platform, they typically underlie the functionality of many of the other input generation approaches and services discussed in this section. Typically these frameworks offer developers and testers an API for writing GUI-level tests for mobile apps through hand-written or recorded scripts.  These scripts typically specify a series of actions that should be performed on different GUI-components (identifying them using varying attributes) and test for some state information via assertion statements.
	
	While useful for developers, these tools are not without their shortcomings. While these frameworks typically provide cross-device compatibility of scripts in \textit{most} cases, there may be edge cases (\eg differing app states or GUI attributes) where scripts fail, highlighting the \textit{fragmentation} problem.  Also, they typically support only a \textit{single testing objective}, as few tools offer support for complex user actions such as scrolling, pinching, or zooming or interfaces to simulate contextual states, which is required for effectively carrying out complex testing scenarios.  More problematic, however, is that GUI level tests utilizing these frameworks are \textit{very} expensive to maintain as an app evolves, discouraging many developers from adopting them in the first place.  	 

%-----------------------------------------------------------------
\subsection{Record and Replay Tools}
\label{subsec:record-replay}

	Manually writing test scripts for mobile GUI or system tests can be tedious, time consuming, and error prone. For this reason both academic and industrial solutions for \textit{Record \& Replay (R\&R)} based testing have been devised. R\&R is an attractive alternative to manually writing test scripts from an ease of use viewpoint, as it enables testers with very limited testing knowledge to create meaningful test scripts for apps.  Additionally, some of the R\&R approaches offer very fine grained (e.g., millisecond accuracy) capture and replay of complex user actions, which can lend themselves well to testing scenarios which require such accuracy (\eg deterministically testing games) or portions of apps that require fined grained user input (\eg panning over a photo or a map).
	
	However, despite the advantages and ease of use these types of tools afford, they exhibit several limitations. Most of these tools suffer from a trade-off between the timing and accuracy of the recorded events and the representative power and portability of recorded scripts.   For context, some R\&R-based approaches leverage the \texttt{/dev/input/event} stream situated in the linux kernel that underlies Android devices.  While this allows for extremely accurate R\&R, the scripts are usually coupled to screen dimensions and are agnostic to the actual GUI-components with which the script interacts. On the other hand, other R\&R approaches may use higher-level representations of user actions, such as information regarding the GUI-components upon which a user acts.  While this type of approach may offer more flexibility in easily recording test cases, it is limited in the accuracy and timing of events.  An ideal R\&R approach would offer the best of both extremes, both highly accurate and portable scripts, suitable for recording test cases or collecting crowdsourced data.  R\&R requires oracles that need to be defined by developers, by manually inserting assertions in the recorded scripts or using tool wizards.
	
%-----------------------------------------------------------------
%Comparison Table of Android Testing Approaches
% !TEX root = ../main.tex
\begin{table*}[p]
\begin{centering}
\caption{This Table Surveys the current state of tools, frameworks, and services that support activities related to mobile testing, originating from both Academic and Industrial backgrounds.}
\vspace{-0.3cm}
\label{tab:android-testing-overview}
\makebox[1\linewidth][c]{       %centering table
\resizebox{1 \linewidth}{!}{
\begin{tabular}{|l|c|c|c|c|c|c|c|}
\hline
\rowcolor[HTML]{A7C6DA} 
\multicolumn{8}{|c|}{\cellcolor[HTML]{A7C6DA}{\ul \textbf{Automation Frameworks \& APIs}}}                                                                                                                                                                                                                                                                                                                                                                                                                                                                                                                                                                                \\ \hline
\rowcolor[HTML]{A7C6DA} 
{\ul \textbf{Name}}                                                    & {\ul \textbf{GUI-Automation}}                                                    & {\ul \textbf{OS API Automation}}                                                 & {\ul \textbf{Black Box}}                                                              & {\ul \textbf{Test-Case Recording}}                                                & {\ul \textbf{Cross-Device Support}}                                               & {\ul \textbf{Natural Language Test Cases}}                                       & {\ul \textbf{Open Source}}                                              \\
\hline
\end{tabular}}}
\end{centering}
\vspace{0.1cm}
\end{table*}
%-----------------------------------------------------------------

%-----------------------------------------------------------------
\subsection{Automated Test Input Generation Techniques}
\label{subsec:test-gen}

	Perhaps the most active area of mobile software testing research has been in the form of the \textit{Automated Input Generation (AIG)} techniques.  The premise behind such techniques is the following: Because manually writing or recording test scripts is a difficult, manual practice, the process of input generation can be automated to dramatically ease the burden on developers and testers.  Such approaches are typically designed with a particular goal, or set of goals in mind, such as achieving high code coverage, uncovering the largest number of bugs, reducing the length of testing scenarios or generating test scenarios that mimic typical use cases of an app.  AIG approaches have generally been classified into three categories\cite{Shauvik:2015,Moran:ICST16}: \textit{random-based} input generation \cite{android-monkey,Machiry:FSE13,Sasnauskas:WODA14,Ravindranath:Mobisys2014}, \textit{systematic} input generation \cite{Amalfitano:ASE12,Anand:FSE12,Azim:OOPSLA13,Moran:ICST16,google-robo-test}, and \textit{model-based} input generation \cite{Amalfitano:IEEE14,Azim:OOPSLA13,Choi:OOPSLA13,Zaeem:ICST2014,Yang:FASE13,Linares:MSR15,Zhang:ICSE17}.  Additionally, other input generation approaches have been explored including search-based and symbolic input generation \cite{Jensen:ISSTA13,Mirzaei:ISSRE15,Mahmood:FSE14,Mao:ISSTA16}.  Nearly all of these approaches can trace their origins back to academic research, with companies like Google just recently entering the market with software-based automated testing services \cite{google-robo-test}.  We provide at-a-glance information about these categories of approaches in Table \ref{tab:android-testing-overview}.  
	
	Research on this topic has made significant progress, particularly in the last few years, however, there are still persistent challenges.  Recent work by Choudhary et. al. \cite{Shauvik:2015} illustrated the relative ineffectiveness of many research tools when comparing program coverage metrics against a naive random approach and highlighted many unsolved challenges including generation of system events, the cost of restarting an app, the need for manually specified inputs for certain complex app interactions, adverse side affects between different runs, a need for reproducible cases, mocking of services and inter-app communication, and a lack of support for cross-device testing scenario generation.  While headway has been made regarding some of these challenges in recent work \cite{Moran:ICST16,Mao:ISSTA16}, many have not been fully addressed.   The specific limitations of these tools again fail to address broader challenges, including flaky tests, fragmentation, limited support for diverse testing goals, and inadequate developer feedback mechanisms.

\vspace{-0.2cm}
%-----------------------------------------------------------------
\subsection{Bug and Error Reporting/Monitoring Tools}
\label{subsec:bug-reporting}

	These types of tools have grown to become an integral part of many mobile testing workflows.  There are two types of tools in this category: (i) tools for supporting bug reporting (a.k.a., issue trackers), and (ii) tools for monitoring crashes and resource consumption at run-time (e.g., New relic\cite{NewRelic} and Crashlytics~\cite{Crashlytics}). Classic issue trackers only allow reporters to describe the bugs using textual reports and by posting additional files such as screenshots; but, real users can only report the bugs when an issue tracker is available for the app, as is the case of open source apps. In the case of tools for monitoring, if developers do not choose to include third-party error monitoring in their application (or employ a crowd-based approach), typically, the only user-feedback or in-field bug reports they receive are from user reviews or limited automated crash reports.  Unfortunately, many user reviews or stack traces without context are unhelpful to developers, as they do not adequately describe issues with an application to the point where the problem can be reproduced and fixed.  In order to mitigate these issues regarding visibility into application bugs and errors, several tools and services exist that aim to help developers overcome this problem.  These tools typically employ features that give developers more detailed information, such as videos \cite{testfairy,appsee,bugclipper,watchsend} or test scripts \cite{Moran:MobileSOFT17}, on failures with concrete reproduction steps or stack traces	 (e.g., crashes); however, to collect that information, the apps need to include API calls to the methods provided by the services. Additionally, they may provide analytic information about how users typically interact with an app, or assist end-users in constructing useful bug reports for apps \cite{Moran:FSE15,Moran:ICSE16,Moran:FSESRC2015}.  Unfortunately, the automated error monitoring tools  are limited to crash reporting (i.e., exceptions and crashes), restricting their utility.	

\vspace{-0.2cm}
%-----------------------------------------------------------------
\subsection{Mobile Testing Services}
\label{subsec:services}

	  Due to the sheer number of different technical challenges associated with automated input generation, and the typically high time-cost of manually writing or recording test scripts for mobile apps, \textit{Mobile Testing Services} have become a popular alternative that utilize groups of human testing experts, or more general crowd-based workers.  This allows the cost of test case generation or bug finding to be amortized across a larger group of workers compensated for their time devoted to testing.  There are typically four different types of testing services offered including: (i) Traditional \textit{Crowd-Sourced Functional Testing} \cite{pay4bugs,testarmy,crowdsourcedtesting,crowdsprint,mycrowdqa,99tests,applause,testio,userlytics,testflight,swrve,loop11,azetone,userzoom,lookback,apptimize} which employs both experts and non-experts from around the world to submit bug reports relating to problems in apps, and who are compensated for the number of \textit{true} bugs that are uncovered; (ii) \textit{Usability testing} \cite{crowdsourcedtesting,testarmy,crowdsprint,mycrowdqa,userzoom,99tests,testio,loop11,azetone,apptimize} aims to the measure the UX/UI design of an app with a focus on ease of use and intuitiveness; (iii) \textit{Security Testing} \cite{testarmy,applause,crowdsprint,99tests,apperian}, which aims to uncover any design flaws in an app that might compromise user security, and (iv) \textit{Localization Testing} \cite{crowdsourcedtesting,mycrowdqa,99tests,applause}, which aims to ensure that an app will function properly in different geographic regions with different languages across the world. While these services do partially address some of the broader challenges of mobile testing such as fragmentation and support for limited testing goals, there are still several notable remaining challenges.  None of these frameworks are open source or free, restricting developers from freely collecting critical usage data from the field which could improve general challenges such as test flakiness or history agnosticism by modeling collected information. Additionally, due to the time cost required of such crowdsourced services, they are typically not scalable in agile development scenarios where an app is constantly changing and released to customers.

%-----------------------------------------------------------------
\subsection{Device Streaming Tools}
\label{subsec:streaming}

	Tools for \textit{Device Streaming} can facilitate the mobile testing process by allowing a developer to mirror a connected device to their personal PC, or access devices remotely over the internet.  These tools can support use cases such as streaming secured devices to crowdsourced beta testers, or providing Q/A teams with access to a private set of physical or virtual devices hosted on company premises.  They range in capabilities from allowing a connected device to be streamed to a local PC (Vysor \cite{vysor}) to open source frameworks and paid services that can stream devices over the internet with low-latency (OpenSTF \cite{stf} \& Appetize.io \cite{appetize}). These tools, particularly OpenSTF, could support a wide range of important research topics that rely on collecting user data during controlled studies or investigations or tools related to crowdsourced testing.

\section{Challenges for Enabling CEL Mobile Testing}
\label{sec:challenges}
% !TEX root = main.tex

	Despite the plethora of methods, techniques, and tools available for automated testing, manual testing of mobile apps is preferred over automated testing due to several factors including personal preferences, organizational restrictions, and current tools lacking important functionality \cite{Mona:ESEM13,Linares-Vasquez:ICSME15,Kochhar:ICST15}. These factors are rooted in a set of challenges that makes mobile app testing a complex and time-consuming task, stemming from both the \textit{inherent properties of mobile smart devices} as well as \textit{unique pressures and constraints on mobile development teams}.  Compared to desktop or web applications, mobile apps are highly event-driven, accepting inputs from users and environmental context shifts, facilitated by a diverse set of sensors and hardware components that provide interfaces for touch-based gestures, temperature measurements, GPS locations, orientation, etc. These diverse input scenarios are difficult for developers to emulate in controlled testing environments. Furthermore, time and budget constraints related to testing practices of startups, small development teams, and even large companies, reduce the possibility for testing apps against a large set of  configurations that are representative of ``in-the-wild" conditions.  

	These two overarching themes, namely inherent technical challenges associated with modern mobile platforms and unique development constraints, underlie a broad set of \textit{key challenges} that we identify within the context of this paper.  In this section we describe a set of distinct open problems,  serving as the most prominent deterrents to enabling a comprehensive automated mobile testing solution, derived from our research experience, industrial collaborations, and conversations with industry professionals.  More specifically, these ``challenges" refer to (i) burdensome components of the mobile testing process that directly (and drastically) impact the testing process,  (ii) open problems for which there are no automated solutions (yet), and (iii) facets of the mobile testing process that have not yet been fully investigated by research or industry.

%-----------------------------------------------------------------
\subsection{Fragmentation}
\label{subsec:fragmentation}

	Mobile app marketplaces allow developers to reach an unprecedented number of users through their online stores. Therefore, in order for an app to be successful, mobile developers must assure the proper functioning of their apps on nontrivial sets of configurations due to the diversity of existing devices.  These sets of configurations can be represented as a testing matrix combining several variations of OSes, versions, and devices. For instance, Nexus 4 devices were originally shipped with Android KitKat, however, as over-the-air updates were pushed to the device, different OS versions were installed by the user-base, a common pattern driving complexity. 
	
	This phenomena is known as \emph{fragmentation} and has been widely recognized as one of the largest challenges when testing apps. Fragmentation is more notable in the case of Android, because the open nature of the platform has led to a large number of devices with different configurations available to consumers \cite{Han:WCRE2012,Mona:ESEM13,Nagappan:FSE14}. As of the time of writing, 25 versions of the Android OS have been released with 7 of the 25 releases (4.1, 4.2, 4.3, 4.4, 5.0, 5.1, 6)\cite{Android-dashboard} accounting for about 97\% of the marketshare across a multitude of hardware. In the case of iOS, the fragmentation is lower, as the number of devices is limited and controlled by Apple, and the market share is more biased toward devices with the latest OS. According to the Apple Store\cite{appstore-dashboard}, as measured on February 2017, 95\% of the iOS users are concentrated in only two versions of the OS (iOS 9 and iOS 10). Despite marketshare consolidation of iOS versions, regular updates to devices create a non-trivial test matrix for developers with 39 distinct combinations when considering only iOS 9 and iOS 10\cite{iosmatrix}. 

	As outlined in the previous section, one potential solution to this fragmentation issue, is cloud/crowd-based services that provide developers with the ability to test their apps on farms of virtual/physical devices, or with a crowd of users with different devices. However, available cloud/crowd-based services are paid, which restrict the type of companies/teams/labs that can afford the services.  In addition, the time typically required to carry out this type of testing is not amenable to agile and DevOps practices.

%-----------------------------------------------------------------
\subsection{Test Flakiness}
\label{subsec:test-flakiness}

	Many modern mobile apps heavily rely on back-end services to (i) delegate operations that can not be performed on the device (e.g., lengthy computations and storage), and (ii) access third-party services such as authentication and geo-localization \cite{Abolfazli:IEEECST}. The bigger the dependency on back-end servers/services, the more prone an app is to suffer from race-condition scenarios or impacted by loss of service due to a lack of connectivity, response time outs, and data-integrity issues when large amounts of data are transferred to and from a device. These conditions can introduce non-determinism in app behavior and outcomes, directly impacting testing: test cases can fail (or succeed) due non-deterministic outcomes affecting assertions. This phenomenon is known as test flakiness \cite{Luo:FSE14}.

	Flaky tests typically emerge when testing complex apps due to assumptions codified in test scripts and test sequences (e.g., inter-arrival delay between events), and lack of mechanisms for identifying (in the app and the tests) unexpected scenarios when invoking back-end servers/services. Other sources of ``flakiness" are varying runtime device states that depend on the available resources when executing the tests. For instance, GC (Garbage Collection) events (in Android apps) can be randomly triggered when the device is running out of memory, and these events can induce GUI lag and even Application Not Responding (ANR) errors \cite{Liu:ICSE14,Linares-Vasquez:ICSME15}. Automation API/Frameworks such as Espresso execute GUI events in test cases only when the GUI is idle, which reduces the risk of flaky tests because of longer-than-expected inter-arrival delays. However, test flakiness is still a major challenge for automated approaches, both from the GUI front-end, and services back-end points of view.

%-----------------------------------------------------------------

\subsection{Mobile-Specific Fault Model and its Application}
\label{subsec:fault-model}

	A key component for measuring test effectiveness or automatically generating effective tests for any software platform is a robust understanding of the faults typically associated with that platform. Mobile apps utilize a programming model different from other more ``traditional" platforms like the web mainly due to mobile-specific factors such as contextual events, event-driven behavior, and gestures-driven interaction.  Previous efforts from researchers and practitioners have been devoted to build catalogs of faults (\aka \textit{fault models}, or \textit{fault profiles}) that describe recurrent (or uncommon) bugs associated with software developed for a particular platform or domain~\cite{Alexander:2002,Gordeyev:2008,Offutt:ISSRE01,Baekken:2006}.  Derived profiles are platform specific and can be used for designing/deriving test cases and for simulating faults during mutation testing~\cite{Ma:ISSRE03,KimCM0,Moeller93,Ostrand:2002:DFL:566171.566181,Zhang08,Nistor:2013:DRF:2487085.2487134, Robinson2009}. In the case of mobile apps, a few works have analyzed/produced bug catalogs mainly focused on performance and energy-related~ issues\cite{Pathak:HOTNETS11,Pathak:Mobisys12,Guo:ASE13,Liu:ICSE14,Linares-Vasquez:ICSME15}.  A recent work by  Linares-V\'asquez \etal \cite{Linares-Vasquez:FSE17} reports a taxonomy of bugs in Android apps derived from manual analysis of different sources such as bugs reported in papers, issue trackers, commit messages, and StackOverflow discussiones. One of the main conclusions in the study is that object-oriented fault models only represent/describe a \textit{subset} of the faults and bugs in mobile apps \cite{Linares-Vasquez:FSE17}. As of today there is no specific usage of mobile-specific fault models for deriving, manually or automatically, test cases or mutation operators. Fault models for mobile apps can also help to statically detect issues before publishing the apps in the markets. However, none of the current approaches for automated testing of mobile apps (described in \secref{sec:state-art}) uses fault models.

%-----------------------------------------------------------------
\subsection{Lack of History Awareness in Test Cases}
\label{subsec:history-aware}

	System and acceptance testing aim to exercise an application under test by simulating user behavior with functional scenarios combining different use cases. For example, testing the correct behavior when deleting or updating a task in a TODO list, first requires the creation of the task. Designing test cases, for both system and acceptance testing, involves designing event sequences that have ``memory" and are aware of the features and uses cases exercised previously.

	History awareness in test cases for mobile apps has been primarily explored in two ways using: (i) models constructed via event-flow graphs (EFG)~\cite{Amalfitano:IEEE14,Azim:OOPSLA13,Choi:OOPSLA13,Zaeem:ICST2014,Yang:FASE13}, or (ii)  language models \cite{Tonella:ICSE14,Linares:MSR15}.  EFGs model the GUI behavior as a finite state machine consisting of various GUI states (\eg windows/screens) with transitions (between the states) defined as input events (\eg click on OK button); EFGs are ripped automatically from the app during runtime, derived a-priori manually or by using static analysis tools like GATOR \cite{Rountev:CGO14,Yang:ICSE15}, or by some combination of these approaches. Language models are probabilistic distributions computed over sequences of tokens/words; in the case of testing, the words are GUI events, and the distributions are generated by analyzing execution traces collected during manual execution of the app. 

	While EFGs and language models are a first step towards history aware models, the derived sequences are prone to be invalid. EFGs do not have explicit memory, but history aware executions can be derived as an artifact of traversing the EFGs. In the case of language models, the memory is explicitly modeled with conditional probability distributions that generate the next event in a test sequence based on its probability conditioned to the occurrence of the previous events \cite{Tonella:ICSE14,Linares:MSR15}. Both, EFGs and language models are not able to recognize high-level features (\ie uses cases) in test cases;  the models generate event sequences but without recognizing what features/use cases the events belong to, thus, the generated test cases are sequences of events instead of sequences of use cases.  Furthermore, the previously described test-flakiness problem makes proper history-aware test case generation problematic.

%-----------------------------------------------------------------
\subsection{Difficulties Evolving and Maintaining GUI Scripts/Models}
\label{subsec:script-evolution}

	Of the available approaches for automated testing of mobile apps, test scripts recorded manually or written with automation APIs/Frameworks are the most vulnerable to app evolution and fragmentation~\cite{Mona:ESEM13,Shauvik:2015,Kochhar:ICST15}. Generating test scripts is time consuming because it requires a practitioner to record/write the test for each target device. However,  these scripts are typically either coupled to display locations (e.g., RERAN), or impacted by reactive GUI-layout changes across devices with differing dimensions. As an application evolves, test scripts need to be updated when changes modify the GUI (or GUI behavior) as expected in the scripts (\eg, the id of a component is modified or a component is removed from a window). Automation APIs like Espresso allow for declaring GUI events partially decoupled from device characteristics, but, the scripts are coupled to change-prone component ids. As of today there is no current approach for automatically evolving scripts written or recorded using Automation APIs.

	One potential solution to this problem is to employ models, as some AIG approaches do.   In theory, a model could be co-evolved as an app changes, and test cases could be generated from the continuously updated model.  While this could currently be accomplished by automated GUI-ripping approaches (and tools based on systematic exploration) that extract the model at runtime, because the model is generated ``just-in-time", this approach wastes the potentially useful knowledge embedded in previously generated models. Therefore, there is a need for techniques that implement intelligent continuous model evolution for the purpose of test case generation.

%-----------------------------------------------------------------
\subsection{Absence of Mobile-Specific Testing Oracles}
\label{subsec:oracles}

An oracle is a set of conditions or expected results that define when a test case has ``failed".  Current tools and practices heavily rely on  developers and testers to (i) manually verify expected results, and (ii) to manually codify oracles via assertions or exceptions when using Automation APIs~\cite{Mona:ESEM13,Kochhar:ICST15,Linares-Vasquez:ICSME15}. In addition to \emph{Manually-Codified-Oracles (MCO)}, other types have been explored, including: 

\begin{itemize}[leftmargin=*]
\item \emph{Exceptions-As-Oracle (EAO)}:  technically, there is no predefined oracle, however, crashes and errors that are reported during the execution of a test case \cite{Hu:AST11,Shauvik:2015,Moran:ICST16,Hu:EuroSys14,Malek:SEREC12,Mahmood:AST12,Mao:ISSTA16,Linares:MSR15,Zaeem:ICST2014,Lin:TSE14} are considered as the oracles for determining whether the case has passed or failed;
\item \emph{GUI-state-As-Oracle (GAO)}: expected transitions between windows and GUI screenshots have been proposed as potential solutions for automatic detection of GUI errors in Android apps ~\cite{Lin:TSE14}. Thus, this type of approach requires (i) executing the test suite on an app version that is considered as the baseline for the testing process, and (ii) collecting the oracles (e.g., screenshots or GUI hierarchy snapshots after each step in the test case). The oracles are then used to verify GUI states in later tests, under the assumption that the GUI-states will remain the same.  
\end{itemize}

	While some promising approaches have been proposed in the context of AIG approaches \cite{Lin:TSE14,Zaeem:ICST2014}, automatic generation of \textit{robust} test oracles is still an open issue. Most of the automated approaches that rely on EAO, are useful for detecting crashes and unexpected errors, but lack capabilities for identifying errors occurring in the GUI; \eg a GUI component expected to be in a window was not activated due to a bug in presentation or business logic where no exception was thrown (or the exception was encapsulated or handled with an empty catch). In addition, using APIs for automated testing (\eg Espresso, Robotium) is very convenient in terms of automatic execution but is also expensive in the sense that they require a developer/tester to write and maintain the tests and corresponding oracle(s). 

	Some automated approaches have proposed a reliance on GAOs implemented as screenshot comparison (\ie expected GUIs vs. resulting GUIs) which get closer to the goal of detecting such GUI-level errors, however, they require image similarity thresholds, which could vary widely for different apps and analyzed windows. The GUI comparison and similarity threshold definition can be impacted by several conditions exhibited under different devices and settings~\cite{Lin:TSE14}. In our experience we have found that the accuracy of matching GUI oracles (i.e., using images as test oracles) can be negatively affected when comparing images (i.e., a GUI oracle against collected GUIs during tests) of different screen sizes and when the test is performed on a different device or orientation (e.g., horizontal vs. vertical). Additionally, GUI hierarchies are typically rendered differently (e.g., hiding/showing some components) on smaller displays because of responsive design capabilities provided by the mobile frameworks or defined programmatically by the developer, which can cause further issues in matching screenshot pairs on devices with diverse configurations.  Further issues with GAOs are: differing color schemes and internationalization. Color palettes can be modified drastically in a target device according to the theme defined by the user or the language setting, which can cause false positives when using images as test oracles. Moreover, GUI components displaying dynamic content, such as date/time pickers or map visualizations, can differ based on context. For instance, collecting a GUI oracle on a \texttt{CalendarView} in an app on February 29\textsuperscript{th}, and then running the test cases ten days later could result in a failing test. 
\vspace{-0.1cm}
%-----------------------------------------------------------------
\subsection{Missing Support for Multi-Goal Automated Testing}
\label{subsec:multi-goal}

	Thus far, AIG approaches have typically centered around \textit{destructive-testing} \cite{destructive-testing}, aimed at eliciting failures from mobile apps.  However, this is only a small subset of the different types of testing required by mobile developers to ensure properly functioning, high-quality apps.  Other types of important testing include: regression, functional, security, localization, energy, performance and play testing.  While some recent work has explored energy-aware test prioritization \cite{Jabbarvand:ISSTA16}, very few other testing goals have been considered by AIG approaches.  
	
	Arguably, other types of testing, particularly functional and regression testing, carry more value than destructive testing as they ensure that an app is functioning \textit{according to its design and requirements.}   Generating tests for functional and regression testing goals is a difficult problem, as an effective approach is required to draw logical links between high-level requirements or use cases and event sequences that would properly exercise and validate these use cases.  Drawing links across abstraction levels in this manner is a challenging task.  In terms of security testing, while approaches for identification of mobile malware have been an active topic of research \cite{Avdiienko:ICSE15,Gorla:ICSE14} test case generation, or automated identification of security vulnerabilities can still be made practical and more robust.  Other than studies into the topic \cite{Linares-Vasquez:ICSME15,Liu:ICSE14}, automated approaches addressing performance testing have also not been extensively investigated. Finally, automated game or play-testing is known to be a technically difficult issue.  Game designers/developers test games by hiring users that play the game under development and provide them with useful feedback regarding bugs, feature requests, and any type of enhancement that could be applied as the game; this type of testing is known as play-testing, and as of today is the only effective way for testing games, given (i) the game interaction through special devices (\eg joysticks, remote controllers, and sensors) or gestures on tactile displays (\eg Wii U Gamepad); and (ii) the need for high-level cognition tasks such as reading, strategic thinking, and visual patterns recognition, that are required when playing games.

\section{Vision: How to Enable CEL Mobile Testing}
\label{sec:vision}
% !TEX root = main.tex 

	To instantiate a coprehensive mobile testing solution, the challenges listed in \secref{sec:challenges} \textit{must} be addressed. The CEL testing framework is based on three core principles aimed at addressing these challeneges: \emph{Continuous}, \emph{Evolutionary}, and \emph{Large-scale}. These principles  integrate and extend concepts from software  evolution and maintenance, testing, agile development, and continuous integration. However, the principles alone are not enough to provide solutions to the aforementioned challenges. Therefore, as part of the CEL testing vision, we propose a system architecture for automated mobile testing following CEL principles. To make this vision tractable, we propose a research agenda for enabling CEL testing and implementing our envisioned system. 
 
 \subsection{The CEL Testing Principles}
\label{vision-principles}
% !TEX root = main.tex

Automated testing of mobile apps should help developers increase software quality within the following constraints: (i) restricted time/budget for testing, (ii) needs for diverse types of testing, and (iv) pressure from users for continuous delivery. Following the CEL principles can enable effective automated testing within these requirements:

\noindent \textbf{Continuous.} Following the principles that support continuous integration and delivery (CI/CD), mobile apps should be continuously tested according to different goals and under different environmental conditions.  Tests should simulate real usages and consider scenarios that simulate different contextual eventualities (\eg exploring a photos app when loosing connectivity) as dictated by app features and use cases. Any change to the source code or environment (\ie usage patterns, APIs, and devices) should trigger --- automatically -- a testing iteration on the current version of the app. To avoid time-consuming regressions, test cases executed during the iteration should cover only the impact set of the changes that triggered the iteration. Finally, to support practitioners when fixing bugs, the bug reports generated with CEL testing should be expressive and reproducible, \ie the bug reports should contain details of the settings, reproduction steps, oracle, inputs (GUI and contextual events), and stack traces (for crashes).
	
\begin{figure*}[ht]
\begin{center}
\vspace{-1cm}
\includegraphics[width=0.85\linewidth]{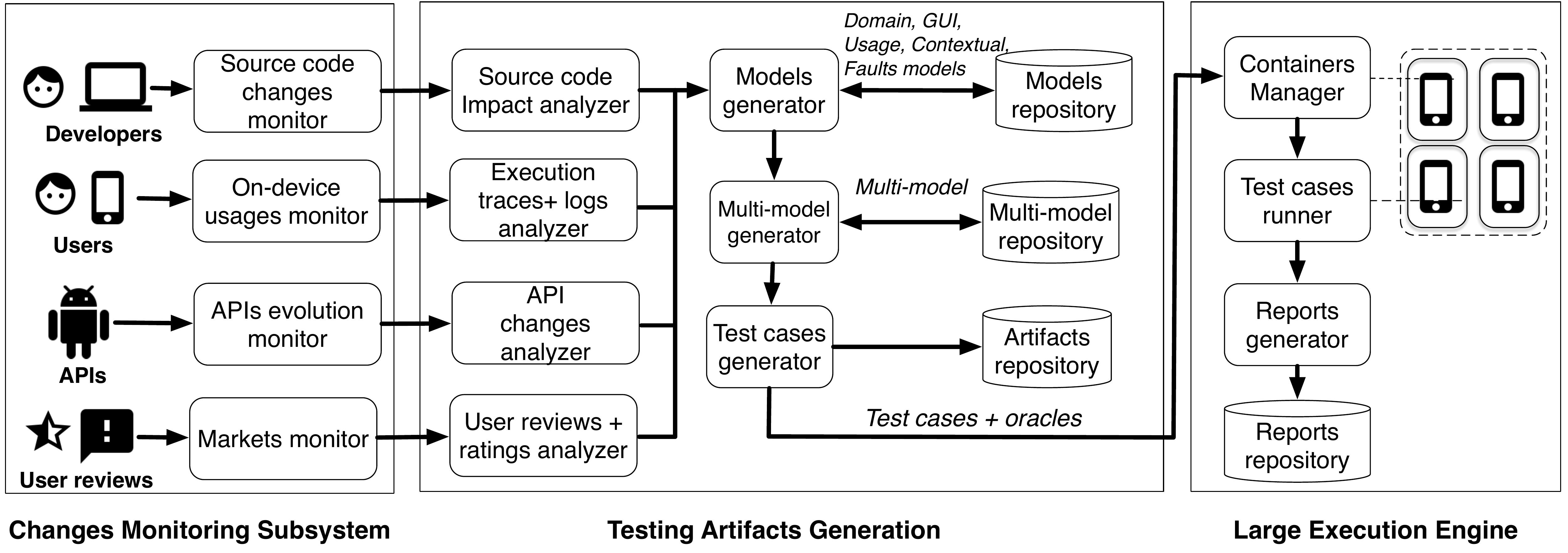}
\vspace{-0.2cm}
\caption{Proposed architecture for a mobile apps testing solution implementing the CEL principles}
\label{vision:arch}
\end{center}
\vspace{-0.7cm}
\end{figure*}
	
\noindent\textbf{Evolutionary.} App source code and testing artifacts (\ie the models,  the test cases, and the oracles) should not evolve independently of one another; the testing artifacts should adapt automatically to changes in (i) the app, (ii) the usage patterns,  and (iii) the available devices/OSes. Thus, the testing artifacts should continuously and automatically evolve, relying not only on source code changes as an input, but also information collected via MSR techniques from sources such as on-device reporting/monitoring, user reviews, and API evolution. CEL testing employs a multi-model representation of the app and this mined data, consisting of GUI, domain, usage, fault, and contextual models, to properly evolve the testing artifacts. This multi-model representation can be used for the evolutionary generation of testing artifacts which consider \textit{both} historical and current data. 
	
\noindent \textbf{Large-scale.} To assure continuous delivery in the face of  challenges such as fragmentation, constrained development timelines, and large combinations of app inputs from GUI and contextual events, CEL requires a large-scale execution engine. This engine should enable execution of test cases that simulate real conditions in-the-wild. Therefore, to support a large test-matrix, CEL testing should be supported on infrastructures for parallel execution of test cases on physical or virtual devices. While virtual devices reduce the cost of implementing the engine, physical devices (or extremely accurate simulations) are mandatory for performance testing. The large-scale engine should be accessible in the context of both cloud and on-premise hardware. Thus, an open-source implementation of the engine is preferred because CEL testing is targeted for both professional development and SE research.

\subsection{Proposed Architecture}
\label{vision-arch}
% !TEX root = main.tex

In this section we propose a conceptual architecture for a solution that follows the CEL principles. \figref{vision:arch} depicts the architecture composed of the following primary components: (i) a change-monitoring subsystem, (ii) a testing artifact generation module, and (iii) a large-scale execution engine.

\subsubsection{Change-monitoring subsystem} The continuous principle suggests that (i) changes to a system under test or its environment should trigger testing iterations; and (ii) test cases should focus on the impact set of the changes to reduce the execution time of testing iterations. The monitoring subsystem should be integrated directly with any source of change or any available mechanism that describes application usages at runtime. Therefore, a CEL testing system should monitor (i) changes pushed to the source code repository or in underlying APIs, (ii) usage patterns collected with on-device reporting/monitoring, and (iii) bugs/crashes reported via user reviews in mobile markets. The rationale for monitoring API changes, is based on the fact that those changes impact directly the quality of mobile apps as perceived by users, because of the change- and bug-proneness of the APIs \cite{Linares-Vasquez:FSE13,Bavota:TSE15,ICSM13:KIM}. Other sources of information that should be exploited by CEL testing systems are app usage patterns in-the-wild and user reviews; both are a valuable source for identifying  sequences of use cases exercised by the users, and bugs/crashes that are context dependent or very hard to identify with  testing. %Examples of  initial efforts on-device reporting/monitor approaches are the RERAN\cite{Gomez:ICSE13} and Valera tools for low-level events recording\cite{Hu:OOPSLA2015}, the Barista tool for GUI-level events recording\cite{FazziniI:ICST16}, and the ODBR approach \cite{Moran:MobileSOFT17}.

\subsubsection{Testing artifacts generation module}
The monitoring subsystem detects changes to the system and its environment, and monitors app usages at run-time and bugs/crashes reported in-the-wild. Any \textit{significant} changes in monitored data should automatically trigger a testing iteration by (i) mining/extracting the information from the mentioned sources in a suitable format for easy analysis, (ii) identifying the impact set of the events (relevant features and code locations, directly and by propagation), (iii) generating/updating underlying models used for deriving test cases, and (iv) generating test cases that are relevant to an event that triggered the testing iteration. %A very promising area is the automated analysis of user reviews with a plethora of works that categorize user reviews\cite{Panichella:ICSME15,Villarroel:ICSE16} and link reviews to source code changes\cite{Palomba:ICSME15} (because of space limitations we only cite the most representative papers).

The decision to use models within the CEL testing architecture is based on capability to express testing scenarios as high-level tokens decoupled from device characteristics (\eg the GUI level model proposed  at \cite{Linares:MSR15}) and their versatility in effectively representing different aspects of a system (\eg GUI flow or domain entities). However,  model-based testing is an area of research that has traditionally focused on \textit{individual} models instead of \textit{multi-model} representations. Testing an app, often requires understanding its GUI, use cases, and domain entities. In the case of testing, mobile-specific fault models can improve the testing process by identifying common faults and risky places in an app. Finally, mobile app execution depends on contextual events such as network connectivity and geo-location capabilities, therefore, a model to represent these features is also valuable when testing apps\cite{Liang:MobiCom14,Zaeem:ICST2014,Moran:ICST16}. 

A CEL testing system should consider models that can be extracted automatically from the different sources including APIs, app usages, and user reviews. The five GUDCF models (GUI, Usage, Domain, Contextual, Fault) should be combined in a unique Multi-Model (MM) representation capable of  generating test cases for different goals. Note that the MM model, particularly the Usage sub-model, should be history-aware.  The GUDCF+MM models should co-evolve automatically with source code changes, and external factors (\ie API evolution, app usages, user reviews) following the CEL evolutionary principle. Evolutionary models allow for on-demand generation of test cases focused on the current needs of the iteration.  For instance, a small commit with a move-method refactoring should only trigger a regression on the impact set, and a large change to the underlying APIs should trigger a testing iteration for all the features that use the modified APIs. Finally, the artifacts generation module, should automatically derive the oracles for the generated cases. The combination of test cases and oracles enables fully automated testing and error reporting. 

\subsubsection{Large scale execution engine} CEL testing requires executing the tests at large-scale to help mitigate the fragmentation problem and reduce the execution time. Large-scale testing can be achieved with farms of physical or virtual devices, allowing for testing across many hardware combinations. Physical devices can be an issue when scaling, in particular for small teams that do not have the budget for a large number of testing devices. Virtual devices are a potential solution, and in particular the ones based on virtual machines that can be instantiated (off-premise or on-premise) using containers (\eg docker-style images). Current device farm services do not offer the flexibility or customization necessary to achieve our CEL vision.  It is worth noting that the test cases to be executed on the devices should be decoupled from device characteristics to reduce test flakiness induced because of invalid GUI events when changing to larger (or smaller) screens. Finally, in the case of bugs/crashes, reports should be expressive enough to guide bug fixing in an effective way. %This type of bug reports can be generated by following the approach suggested by Moran \etal \cite{Moran:ICST16}.

\vspace{-0.1cm}
\subsection{Research Agenda}
\label{vision-agenda}
% !TEX root = main.tex

\vspace{-0.1cm}
Based on the current frameworks, tools, and services that are available to developers, as well as the limitations and remaining open challenges in the domain of mobile testing, we firmly believe that our vision for \textit{Continuous}, \textit{Evolutionary} and \textit{Large-Scale} mobile testing offers a comprehensive architecture that, if realized, will dramatically improve the testing process.  However, there are still many components of this vision that are yet to be properly explored in the context of research.  Therefore, in order to make our vision for the future of mobile testing tractable, we offer an overview of a research agenda broken down into six major topics. 

%-----------------------------------------------------------------
\subsubsection{Improved Model-Based Representations of Mobile Apps}
\label{ssubsec:improving-models}

	Current approaches for deriving model-based representations of apps are severely lacking a multi-model-based approach that might significantly improve the utility of model-based testing.  However, to this end, there are several unexplored areas requiring further research and investigation.  While model-based representations of mobile GUIs have been widely explored \cite{Amalfitano:IEEE14,Choi:OOPSLA13,Zaeem:ICST2014,Yang:FASE13,Linares:MSR15,Zhang:ICSE17}, researchers should focus on unifying the (often complementary) information which can be extracted from both static and dynamic program analysis techniques. For instance, using static control flow information from a tool like GATOR to guide dynamic GUI-ripping to extract a more complete \textit{GUI model}.  Very little research work has been devoted to deriving \textit{domain models} from applications, however, such models will be crucial for enabling automated tests to exercise complex inputs and behaviors.  Future studies could focus on automatically extracting domain models from source code and data storage models, and by examining common traits between apps that exist in similar categories in app marketplaces in order to derive common event sequences and  GUI-usage patterns.
	
	Given the highly contextualized environment of mobile apps (e.g., varying network and sensor conditions), effective automated testing will require a \textit{contextual model} identifying and quantifying the usages of related APIs within in application. While some recent work has explored such functionality \cite{Moran:ICST16}, this can be made more precise and robust through more advanced static analysis and dynamic techniques that infer potential context values to help drive automated testing.  Very few automated testing approaches for mobile apps consider \textit{usage models} \cite{Linares:MSR15,Mao:ISSTA16} stipulating common functional use cases of an app, expressed as combinations of GUI events. Recent advances in deep-learning based representations may be applicable for appropriately modeling user interactions and high-level features, if properly cast to the problem domain.  
	
	In order to better inform test case generation and properly measure the effectiveness of automated testing, platform specific \textit{fault models} must be empirically derived through observations and codification of open source mobile app issue trackers, and knowledge bases such as Stack Overflow \cite{stack-overflow} or the XDA developer forums \cite{xda-developers}.  Finally, in order for these models to be viable within an evolutionary context, there must exist mechanisms for accurate, history aware model updates.  A continuously evolving model will allow for more robust updates to generated test-related artifacts.  

%-----------------------------------------------------------------
\subsubsection{Goal-Oriented Automated Test Case Generation}
\label{ssubsec:improving-models}

	Current approaches for automated input generation for mobile apps have typically focused on a single type of testing, namely \textit{destructive testing} \cite{destructive-testing} or some derivation thereof.  The effectiveness of such techniques are typically measured code coverage metrics or by the number of failures uncovered.  While this type of testing can help improve the quality of an app, it is one of \textit{many} important testing practices in the mobile domain.  In order to provide developers with a \textit{comprehensive} automated testing solution, researchers must focus on automated test generation for other types of testing aimed at different goals, particularly those measuring mobile-specific quality attributes.  Some of these testing types include security testing, localization testing, energy testing, performance testing and play-testing.  Testing for different goals on mobile platforms \textit{fundamentally differs} from similar testing scenarios for other types of software due to the GUI and event-driven nature of mobile apps, and the fact that GUI tests on devices are currently a necessity (as unit testing misses important features untestable outside of device runtimes) for exercising enough app functionality to achieve effective practices for many of these testing scenarios.  Therefore, the challenge to the research community is to utilize the representation power of the models we describe in this paper to devise techniques for automated test case generation for different \textit{testing goals}.  

%-----------------------------------------------------------------
\subsubsection{Flexible Open Source Solutions for Large Scale and CrowdSourced Testing}
\label{ssubsec:improving-models}

	As mobile markets mature and additional devices are introduced by consumer electronics companies, the mobile fragmentation problem will only be exacerbated.  As previously discussed, cloud-based services offering virtually accessible physical devices and crowdsourced testing are two promising solutions to this issue, however, these solutions are not available to all developers and are not scalable to all testing goals. For instance, it may be difficult to carry out effective energy or security testing on cloud-based devices if such services are not specifically enabled by a cloud provider.  As outlined in our vision, we looked to container and virtual machine technology that has made testing practices scalable in development scenarios like continuous integration (CI).  Thus, it is clear that a robust and highly customizable container or virtualization image of a mobile platform is the most promising long-term, scalable solution for enabling our vision of CEL testing.  Future research in the systems area could focus on improving the viability of promising open source source projects as androidx86 \cite{androidx86} to be used in CI-like development environments, allowing for further customizations and control over attributes such as sensor value mocking and screen size and pixel density.  While valuable, these virtual devices will not be applicable to all types of testing, such as usability testing, or usage information collection which can be used to derive an effective usage model of an app.  Instead, such goals fit the model of crowdsourced testing well.  Unfortunately, no flexible open source solutions to support developers or researchers currently exist, signifying the need for such a platform.  Luckily, there are existing modern open source solutions such as OpenSTF \cite{stf} and ODBR \cite{Moran:MobileSOFT17} that could help facilitate the creation of such a platform. This  platform should allow for easy collection of privacy-aware execution traces and logs, suitable for deriving usage models.

%-----------------------------------------------------------------
\subsubsection{Derivation of Scalable, Precise Automated Oracles}
\label{ssubsec:improving-models}

	To allow viable automated support of a diverse set of testing goals, progress must be made in the form of automatically generated, accurate, and scalable oracles.  It is likely that such oracles will be specific to particular types of testing tasks and require different technological solutions.  Some automated testing approaches have broached this problem and devised simple solutions such as using app agnostic oracles based on screen rotation actions \cite{Zaeem:ICST2014} or GUI screenshots as state-representations \cite{Lin:TSE14}.  However, there are still open problems even with these simple types of oracles, and they are not comprehensive.  Promising directions along this research thread might include mixed GUI representations that utilize both image and textual representations of GUI information to form robust state indications, which could be used as automated oracles.  Additionally, the derivation of mobile platform-specific fault models may help in deriving automated oracles that could test for common problems inherent to mobile apps.

%-----------------------------------------------------------------
\subsubsection{Mining Software Repositories and User Reviews to Drive Testing}
\label{ssubsec:improving-models}

	While many different automated testing solutions for mobile apps have been proposed, they largely ignore information sources which could be invaluable for informing the testing process, namely data mined from software repositories and user reviews.  Information from software repositories for mobile apps could be collected in two ways, which could be combined to maximize the information utility, (i) mining the development history of a single application, and (ii), the development history of collections of open source apps hosted on services like GitHub.  Here lightweight static analysis techniques could be used at scale, whereas more expensive app control flow analysis techniques could be used to provide more detailed code-level information about a single subject app.  Mobile app developers also have an unprecedented feedback mechanism from users in the form of user reviews.  As such there is a growing body of work that has focused on identifying informative reviews \cite{Chen:icse2014,Palomba:ICSME15,Villarroel:ICSE16}, linking these to affected areas of source code \cite{Palomba:ICSME15}, and even recommending code changes \cite{Palomba:ICSE17}.  However, little work has been done to use the information contained within informative reviews to drive different types of testing.  For instance, in the context of functional or regression testing, user reviews could be used to prioritize test cases, or even generate test cases for issues derived from reviews.  

%-----------------------------------------------------------------
\subsubsection{Derivation of Methods to Provide Useful Feedback for Developers}
\label{ssubsec:improving-models}

	In order to make the results of automated testing practices \textit{useful} and \textit{actionable} for developers, researchers must dedicate effort to (i) deriving useful visual representations of testing results, and (ii) augmenting typical methodologies by which users might report feedback to developers.  Very few automated testing approaches have considered methodologies for augmenting or effectively reporting testing information to developers \cite{Moran:FSE15,Moran:ICST16}.  Here researchers might consider applications of promising visualization approaches adopted from the HCI community combined with developer information needs derived from empirical studies. The studies conducted with engineers can help to develop theoretically grounded solutions for providing them with actionable information and augmented context (\eg  sound traceability links back to different parts of application code).	Additionally, novel mechanisms for aiding users in providing actionable feedback to developers will be important to increase the quality of mineable information (\eg on-device bug reporting and monitoring).

\section{Conclusions}
\label{sec:conclusion}
% !TEX root = main.tex

In this paper we presented a comprehensive perspective of automated mobile testing with the purpose of reporting the current tools and challenges, and envisioning the future of the field. After surveying the current state of the art and challenges related to mobile testing, we provided our vision of future automated mobile testing techniques that enable agile development practices. To this, we defined the CEL testing framework that is based on the \emph{Continuous}, \emph{Evolutionary}, and \emph{Large-scale} principles. This framework includes a proposal of an architecture encompassing a fully automated system for mobile app testing, and a research agenda that describes six major topics that should be addressed in the near future to enable fully automated solutions adhering to CEL principles. We hope this paper will serve as a road map for the mobile testing community to design and implement future approaches.

\balance
\bibliographystyle{IEEEtran}
\bibliography{ms}

% that's all folks
\end{document}